\documentclass[twocolumn]{aastex61}
\usepackage{graphicx}

\shorttitle{V704~And \& RX~J2338+431}

\shortauthors{Weil et al.}

\newcommand{\unit}[1]{\, {\rm #1}}

\begin{document}

\title{An Optical Study of Two VY Sculptoris-Type Cataclysmic Binary Stars:\\ V704~And and RX~J2338+431}

\author{Kathryn E.\ Weil}
\affil{Department of Physics \& Astronomy, 6127 Wilder Laboratory, Dartmouth College, Hanover, NH 03755-3528, USA}

\author{John R.\ Thorstensen}
\affil{Department of Physics \& Astronomy, 6127 Wilder Laboratory, Dartmouth College, Hanover, NH 03755-3528, USA}

\author{Frank Haberl}
\affil{Max-Planck-Institut f{\"u}r extraterrestrische Physik, Giessenbachstra{\ss}e, D-85748 Garching, Germany}

\begin{abstract}

We report observations of the known cataclysmic variable star (CV)
V704~And, and also confirm that the optical counterpart of the ROSAT 
Galactic Plane Survey source RX~J2338+431 is a heretofore-neglected CV.
Photometric and spectroscopic observations from MDM
Observatory show both systems to be novalike variables that exhibit
dips of 4-5 magnitudes from their mean brightnesses, establishing them 
as members of the VY~Scl subclass.  
From high-state emission-line radial velocities, we determine orbital 
periods of 0.151424(3) d (3.63 hr) for V704~And and 0.130400(1) d (3.13 hr) 
for RX~J2338+431. 
In V704 And, we find that the H$\alpha$ emission-line measures 
cluster into distinct regions on a plot of equivalent width versus 
full width at half-maximum, which evidently correspond to high,
intermediate, and low photometric states.  This allows us 
to assign spectra to photometric states when 
contemporaneous photometry is not available, an apparently
novel method that may be useful in studies of other novalikes.
Our low-state spectra of RX~J2338+431 show 
features of an M-type secondary star, from which we estimate a distance of
$890\pm 200$~pc, in good agreement with the Gaia DR2 parallax. 

\end{abstract}


\section{Introduction} Cataclysmic variable stars (CVs) are short-period
binary star systems containing a white dwarf (WD) and a secondary that typically 
resembles a low mass main sequence star. The WD accretes mass as the secondary star
overflows its Roche Lobe. CVs are classified by their photometric and
spectroscopic properties into different subclasses. Those that remain mostly in a high
state, with high mass transfer onto the WD, are classified as nova-like variables
(NLs).  

A subclass of NLs, VY~Sculptoris stars, show occasional rapid drops in
brightness by at least 1.5 magnitudes from their mean
\citep{Warner1995,Hellier2001}.  It is widely supposed
\citet[and references therein]{Hellier2001} that VY~Scl stars change
states due to a variation in the mass transfer rate from the secondary. When the
steady flow characteristic of the high state becomes disrupted, the accretion
disk apparently vanishes and the low state of the system is observed. The cause of
the mass-transfer rate variations is not known, but one proposed
mechanism is the passage of a starspot on the
secondary through inner Lagrangian point, at which the secondary
loses mass. The lower temperatures in the star spot would decrease 
the photospheric scale height, suppressing the mass 
transfer. 

In the low states of VY Scl stars, the broad emission 
lines commonly seen in CVs become weaker and much narrower. Also,
the emission-line radial velocities
in the low state can be up to 180 degrees out of phase with the high-state
velocities \citep{Warner1995}.  For more discussions on the low states of CVs,
see \citet{Schneider1981,Shafter1985,RG2012}. 

In this paper, we present observations of two VY~Scl stars, V704~And and
RX~J2338+431. Table~\ref{tab:objprops} lists their basic properties. 

V704~And, formerly known as LD~317, was first discovered by
\citet{Dahlmark1999} with a measured apparent V magnitude varying between 12.8
and 14.8. 
\citet{Papadaki2006} observed V704~And in 2003
September, 2003 November, and 2005 January and concluded the system had
experienced fading episodes in 2003 and 2005. Our 2003 October observations
showed the system in its high state, indicating the fade to the low state was
rapid. This evolution is also seen in the American Association of Variable
Star Observers (AAVSO) light curves, which show the
star experiencing a fading event in late 2003 October into early 2003 November.
The AAVSO light curves include observations of V704~And dating back to 1966,
and show that there are multiple episodes of rapid drops in brightness ranging
from 1 magnitude up to 4.5 magnitudes. This episodic fading behavior is
consistent with other known VY~Scl stars. 

\begin{deluxetable*}{clllllllllll}[t]
\tabletypesize{\scriptsize}
\tablecaption{Object Properties\label{tab:objprops}}
\tablehead{
\colhead{Object Name}& \colhead{$\alpha$} & \colhead{$\delta$}& \multicolumn{3}{c}{2MASS\tablenotemark{a}}& \multicolumn{5}{c}{SDSS DR12\tablenotemark{b}}& \colhead{ROSAT\tablenotemark{c}}\\
& & & \colhead{J}& \colhead{H}& \colhead{K}& \colhead{g}& \colhead{u$-$g}& \colhead{g$-$i}& \colhead{r$-$i}& \colhead{i$-$z}&\\
& \colhead{[h m s]}&\colhead{[$\arcdeg\ \arcmin\ \arcsec$]} & \colhead{(mag)}& \colhead{(mag)}& \colhead{(mag)}& \colhead{(mag)}& \colhead{(mag)}& \colhead{(mag)}& \colhead{(mag)}& \colhead{(mag)}& \colhead{(counts/sec)} }
\tablecolumns{12}
\startdata
V704~And\tablenotemark{d}  & $23^{\rm h}44^{\rm m}57.5398^{\rm s}$&$+43\arcdeg31\arcmin22.318\arcsec$& 12.892 & 12.714 & 12.579 &\nodata\tablenotemark{e} & \nodata\tablenotemark{e} & \nodata\tablenotemark{e} & \nodata\tablenotemark{e} & \nodata\tablenotemark{e} & $0.03\pm0.01$ \\
RX~J2338+431\tablenotemark{d}  &$23^{\rm h}37^{\rm m}59.2213^{\rm s}$ &$+43\arcdeg08\arcmin50.896\arcsec$ & 15.366 & 15.244 & 14.824 & 16.10 & $-$0.07 & 0.08 & 0.09 & 0.08 & $0.06\pm0.02$\\
\enddata
\tablecomments{}
\tablenotetext{a}{2MASS data from 2MASS All Sky Catalog of point sources \citep{2MASS}.}
\tablenotetext{b}{SDSS data from Data Release 12 \citep{SDSSObs,SDSSFilters}.}
\tablenotetext{c}{Count rates are from the ROSAT Bright Source Catalogue \citep{Voges1996}.}
\tablenotetext{d}{Positions are from {\it Gaia} DR2 \citep{GaiaPaper1,GaiaPaper2,GaiaDR2}.  They
are referred to the ICRS (equivalent to the J2000 equinox).}
\tablenotetext{e}{SDSS observed V704~And, but the data were flagged as unreliable.}
\end{deluxetable*}

\begin{figure*}[htb!] 
  \centering 
  \includegraphics[scale=0.35]{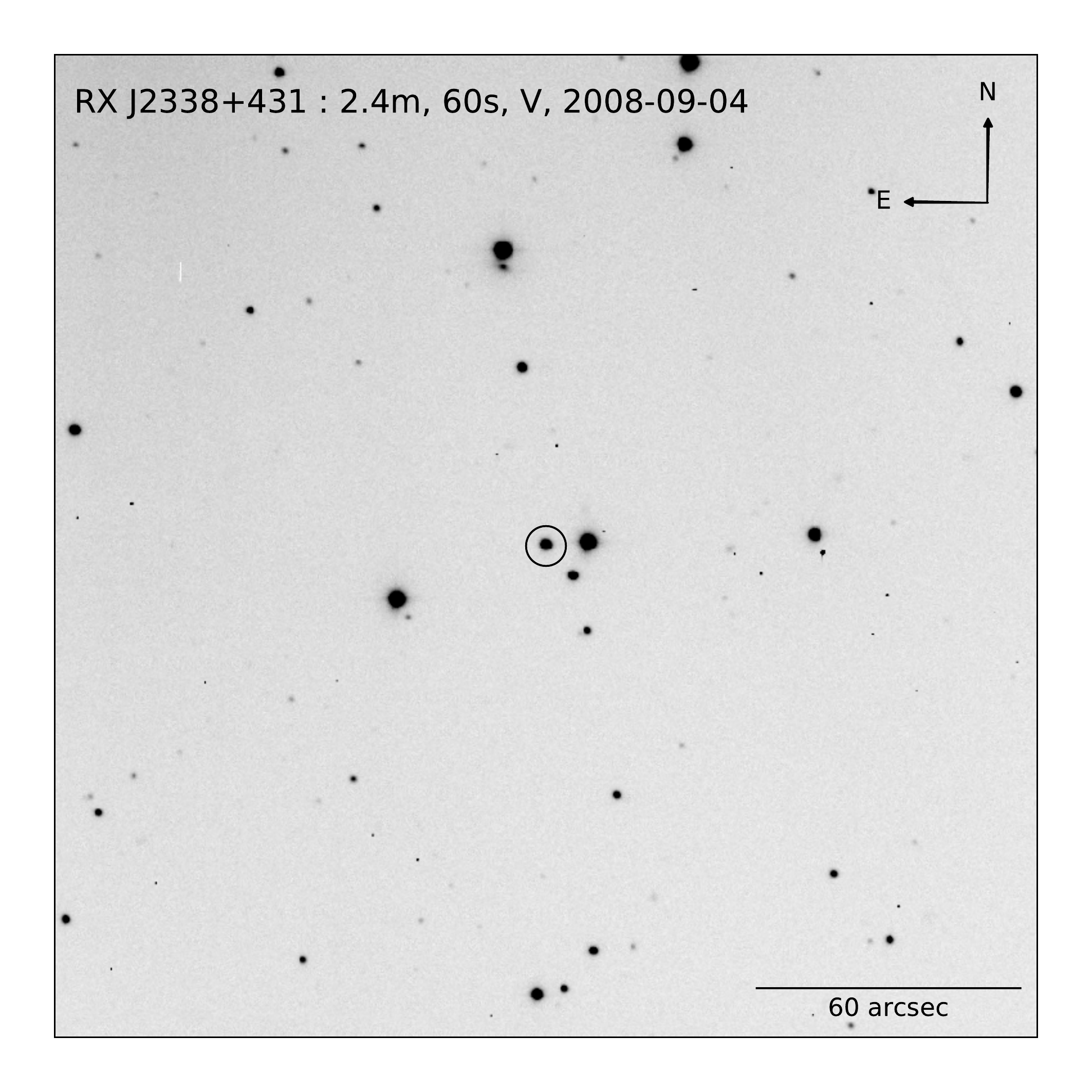}
\caption{V-band exposure with the MDM 2.4m telescope from 2008 September 4 UT. The
circle indicates the cataclysmic.  North is at the top and
east to the left, and the scale is indicated in the bottom right.}
\label{fig:rx2338Chart} 
\end{figure*}

Less has been reported on RX~J2338+431 (Table~\ref{tab:objprops}). 
It was selected from the
ROSAT Galactic Plane Survey \citep[RGPS;][]{Motch1991} as a hard X-ray source,
and is listed in the ROSAT Bright Source Catalogue \citep[1RXS~J233801.0+430852;][]
{Voges1996}\footnote{\citet{Truemper1982} gives an overview of the ROSAT mission.}.
Direct imaging in 1999 October revealed an
ultraviolet object at a location consistent with the ROSAT position
(Table~\ref{tab:rx2338Photo}, Figure~\ref{fig:rx2338Chart}). Guided by the
apparent UV excess from direct imaging, we obtained a spectrum of the UV-bright
object on 1999 October 19 UT that showed emission lines consistent with a CV.
We identify this object as the optical counterpart of 1RXS~J233801.0+430852. 
The identification is noted in the online CV catalogs compiled by 
\citet{downes01} (see also \citealt{downes05}).  The identification
is corroborated by the Swift X-ray Point Source (SXPS) catalog
\citep{Evans2014}, which lists a source only 0.6 arcsec from
the optical counterpart.

\begin{deluxetable}{lcccc}[t]
\tablecaption{RX~J2338+431 Photometry \label{tab:rx2338Photo}}
\tabletypesize{\scriptsize}
\tablehead{
\colhead{Date} &
\colhead{V} &
\colhead{U$-$B} &
\colhead{B$-$V} &
\colhead{V$-$R} \\
& 
\colhead{(mag)} & 
\colhead{(mag)} & 
\colhead{(mag)} & 
\colhead{(mag)} 
}
\startdata
October 1999 & 16.95 & $-$1.00 & $-$0.17 & 0.87 \\
January 2000 & 20.07 & $-$1.68 & 0.48 & 2.10 \\
August 2008 & 16.50 & $-$0.96 & $-$0.03 & 0.73 \\
\enddata
\tablecomments{June 2003 observations are not included because only V-band images were taking during that observing run.}
\end{deluxetable}

\section{Observations} 

The spectral observations reported here
are from the 2.4 m Hiltner telescope and the 1.3 m McGraw-Hill telescope at MDM
Observatory on Kitt Peak, Arizona. At both telescopes, nearly all observations
were taken with the ``modspec'' spectrograph with a $1\farcs1$ slit and ``Echelle''
detector, a SITe $2048^2$ CCD giving a dispersion of 2 \AA\ ${\rm pixel}^{-1}$
between 4210 and 7500~\AA, with declining throughput toward the ends
of the spectral range. When Echelle was unavailable a similar $1024^2$ SITe
detector (``Templeton'') was used, which covered from 4660 to 6730~\AA. For a
few spectra we used the Mark III grism spectrograph, which covered 4580 to
6850~\AA\ at 2.3 \AA\ ${\rm pixel}^{-1}$ resolution.  We used an incandescent
lamp for a flat field, derived the 
wavelength calibration from observations of Hg, Ne, and Xe lamps, 
and observed flux standard stars in twilight when the weather was clear. 

Spectral observations for V704~And were taken between 2001 January and 2016
June, and for RX~J2338+431 between 1999 October and 2002
October.  They cover the objects in their high, intermediate, and low states.  
Since all of our observations are from a single site, the sampling has an
inevitable periodicity at $\sim 1$ sidereal day, which can lead to ambiguities 
in the daily cycle count.  To ameliorate this problem, we took observations over 
a wide range of hour angle \citep{Thorstensen2016}. Tables 
\ref{V704Obs} and \ref{RX2338Obs} list the times of observation.  Exposure
were typically 480 s, but varied from 180 to 1200 s.

To reduce the spectra we used procedures based mostly on IRAF tasks; 
\citet{Thorstensen2016} and \citet{ThorstensenHalpern2013} give more detail.

\begin{deluxetable}{lrrr}
\tablecaption{V704~And Radial Velocities \label{V704Obs}}
\tabletypesize{\scriptsize}
\tablehead{
\colhead{Time\tablenotemark{a}} &
\colhead{$r_{\rm rad}$} &
\colhead{$\sigma_v$\tablenotemark{b}} &
\colhead{State\tablenotemark{c}} \\
& 
\colhead{[km s$^{-1}$]} &
\colhead{[km s$^{-1}$]} &
}
\startdata
52084.97272 &    12 & 3 & H  \\
52085.93072 & $-$14 & 4 & H  \\
52086.84160 &    73 & 6 & H  \\
\enddata
\tablenotetext{a}{Barycentric Julian date of mid-integration, minus 2,400,000.}
\tablenotetext{b}{Minimum uncertainty in radial velocity, based on counting statistics alone.} 
\tablenotetext{c}{Photometric state of the system; H = high, M = medium, and L = low.}
\tablecomments{Table 3 is published in its entirety in the machine-readable format.
A portion is shown here for guidance regarding its form and content.}
\end{deluxetable}

\begin{deluxetable}{lrrr}
\tablecaption{RX~J2338+43 Radial Velocities \label{RX2338Obs}}
\tabletypesize{\scriptsize}
\tablehead{
\colhead{Time\tablenotemark{a}} &
\colhead{$r_{\rm rad}$} &
\colhead{$\sigma_v$\tablenotemark{b}} &
\colhead{State\tablenotemark{c}} \\
& 
\colhead{[km s$^{-1}$]} &
\colhead{[km s$^{-1}$]} &
}
\startdata
51470.77543 & $-$60 & 12 & H \\
51470.77948 & $-$52 & 14 & H \\
51470.78637 & $-$68 & 13 & H \\
\enddata
\tablenotetext{a}{Barycentric Julian date of mid-integration, minus 2,400,000.}
\tablenotetext{b}{Minimum uncertainty in radial velocity, based on counting statistics alone.} 
\tablenotetext{c}{Photometric state of the system; H = high, L = low.}
\tablecomments{Table 4 is published in its entirety in the machine-readable format.
A portion is shown here for guidance regarding its form and content.}
\end{deluxetable}

For RX~J2338+431, photometry was collected using the 2.4 m Hiltner telescope
and Echelle CCD during 1999 October, 2000 January, 2003 June, and
2008 August. The magnitudes were calibrated with observations of
\citet{Landolt1992} standards. The error in the magnitude calibration is likely
not to exceed $\sim 0.1$ mags as judged from the scatter in the standard star fits. 
Figure~\ref{fig:V704rx2338LC} shows the light curve, (blue circles), and 
Table~\ref{tab:rx2338Photo} gives the colors. 

We also synthesized $V$ magnitudes from our flux-calibrated spectra,
by using a passband from \citet{bessell90}, and the IRAF {\it sbands}
task, which multiplies the spectrum by the passband, integrates, and
adds an appropriate constant.  We 
often take spectra through thin cloud, and do not apply any
correction for the light lost
on the slit jaws (which varies with the seeing). 
The synthesized magnitudes are therefore 
less reliable and accurate than the filter photometry, which we only took under
apparently clear skies and for which we used software apertures
large enough to include the great majority of the light.
Experience suggests that magnitudes synthesized from an
average of several spectra are usually accurate to 
$\sim 0.3$ mag, when obviously poor spectra
are excluded from the average.

\section{Analysis} 

The emission-line radial velocities of VY Scl stars behave
differently with photometric state, so in order to find the
periods from radial velocities, we needed to create time
series segregated by state.  However, simultaneous photometry was 
not available for all our spectroscopic observations, and 
we were reluctant to rely on the less-accurate synthesized
$V$ magnitudes, so 
we needed to determine the state without referring 
directly to magnitudes.  To this end, we measured the 
equivalent width (EW, taken to be positive for
emission) and full-width at half-maximum
(FHWM, as determined from Gaussian fits) of the H$\alpha$ 
emission line.  We then looked for correlations with the 
photometric states in cases where magnitudes were available.

\begin{figure*}[t]
   \centering
   \includegraphics[scale=0.4]{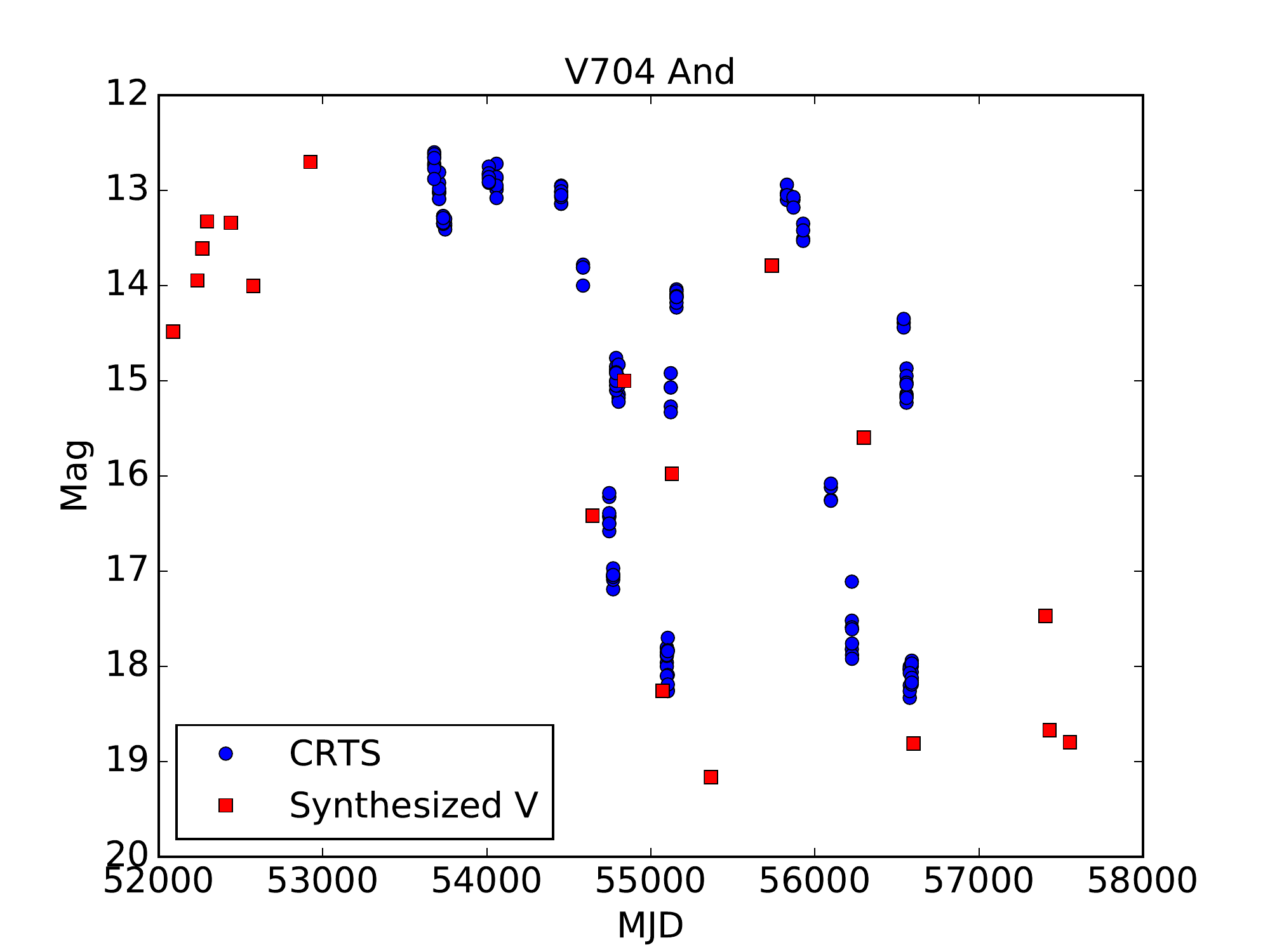}
   \includegraphics[scale=0.4]{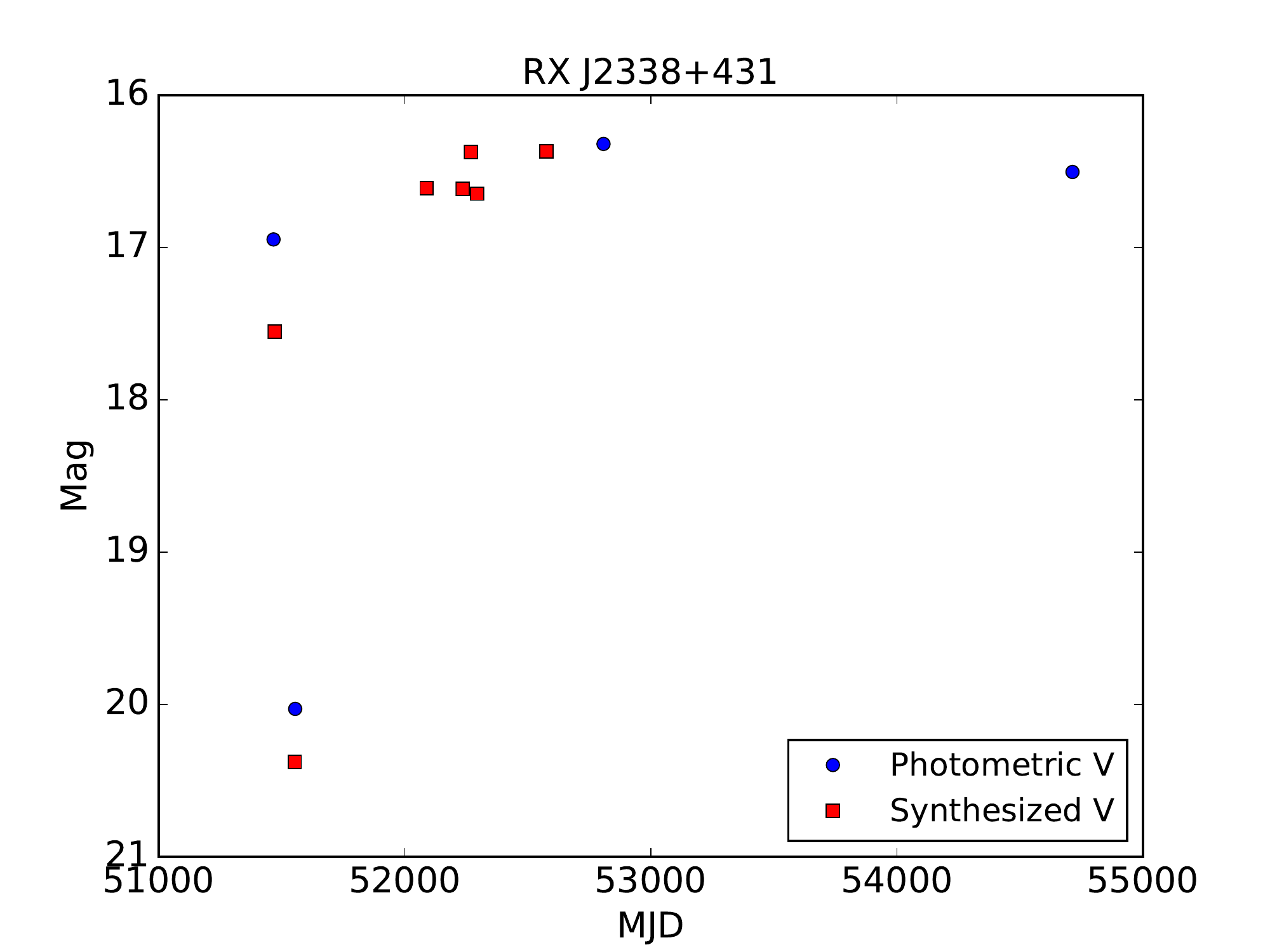} 
   \caption{Light curves for V704~And (left) and RX~J2338+431 (right). The red squares are the synthesized V magnitude determined from the spectral observations taken at MDM observatory. On the left, the CRTS Light Curve for V704~And is shown as blue circles. On the right, blue circles represent the observed V band magnitude from MDM photometric observations. RX~J2338+431 is not covered by CRTS.}
   \label{fig:V704rx2338LC}
\end{figure*}

\begin{figure*}[t]
   \centering
   \includegraphics[scale=0.4]{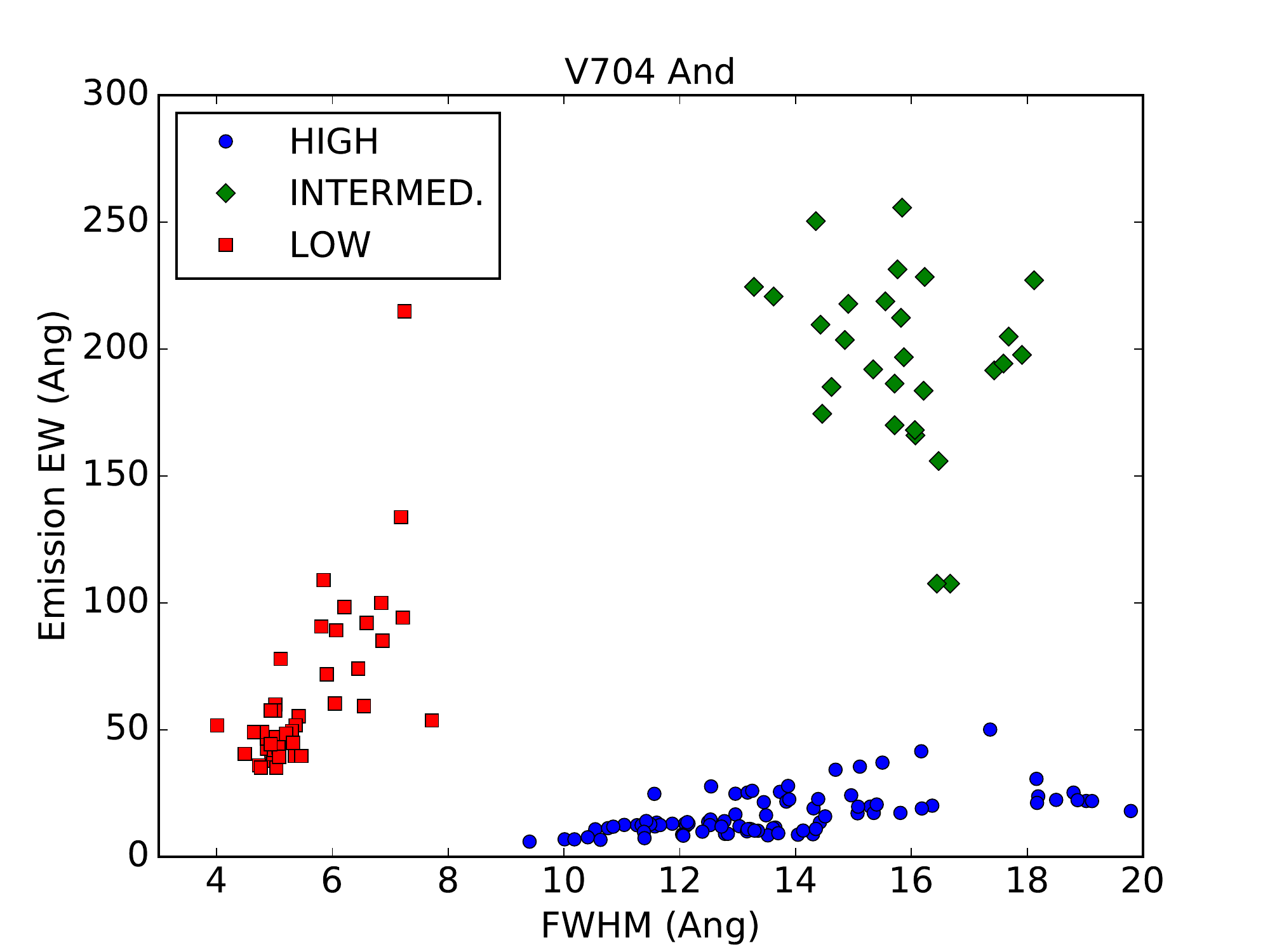} 
   \includegraphics[scale=0.4]{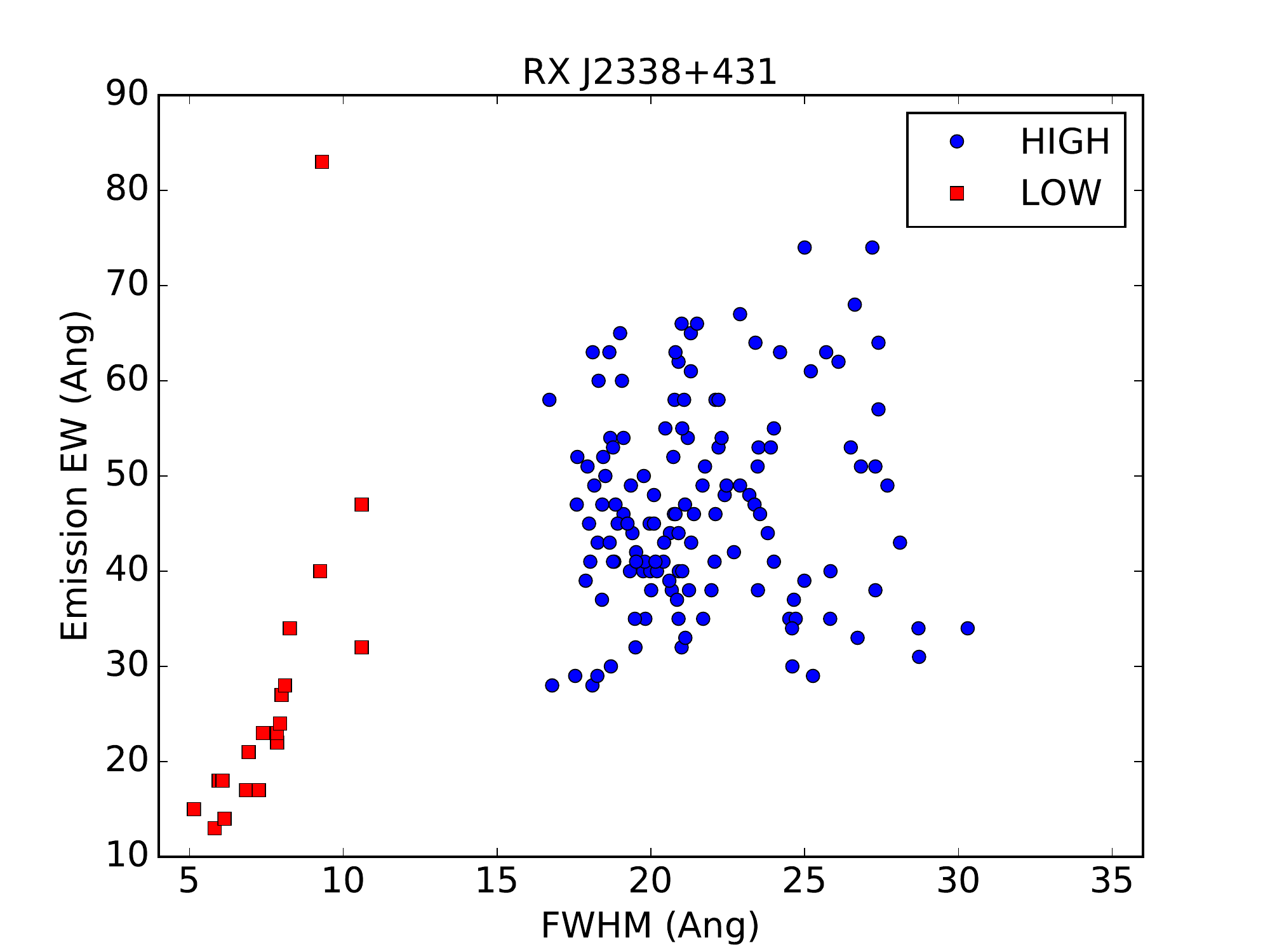} 
   \caption{Emission equivalent width vs. full-width-half-maximum measurements
of the H$\alpha$ emission line, which were used to classify the state of the
CV. \textit{Left:} V704~And, which has measurements in three states: high (blue
circles), intermediate (green diamonds) and low (red squares). \textit{Right:}
RX~J2338+431, which has measurements in two states: high (blue circles) and low
(red squares).}
   \label{fig:V704rx2338EW}
\end{figure*}

Plots of the emission EW vs.~the FWHM showed distinct
groupings that proved to correspond to the photometric state,
and we used these to separate the data into various 
states without reference to the photometric observations. For V704~And, 
three groupings emerged, which can be seen
in Figure~\ref{fig:V704rx2338EW}.  The grouping with the smallest
emission EWs (less than $\approx 51\unit{\AA}$) and FWHMa above $9\unit{\AA}$
(blue circles), corresponds to the photometric high state. 
Spectra in the group with FWHM less than $9\unit{\AA}$ (red
squares) were from the low state, while those with EW above $\approx
51\unit{\AA}$ and FWHM greater than $12\unit{\AA}$ (green diamonds) were
from an intermediate state. For
RX~J2238+431, only two groups emerged, which can be seen in
Figure~\ref{fig:V704rx2338EW}. The high state spectra had the highest FWHM
above $15\unit{\AA}$ (blue circles), while the low state spectra had FWHM less
than $15\unit{\AA}$ (red squares).  As corroboration, we note
that the groupings inferred from the 
EW vs. FWHM analysis corresponded well with the V-band magnitudes 
synthesized from the spectra  (Figure \ref{fig:V704rx2338LC}).

\begin{figure*}[t]
   \centering
   \includegraphics[width=0.75\textwidth]{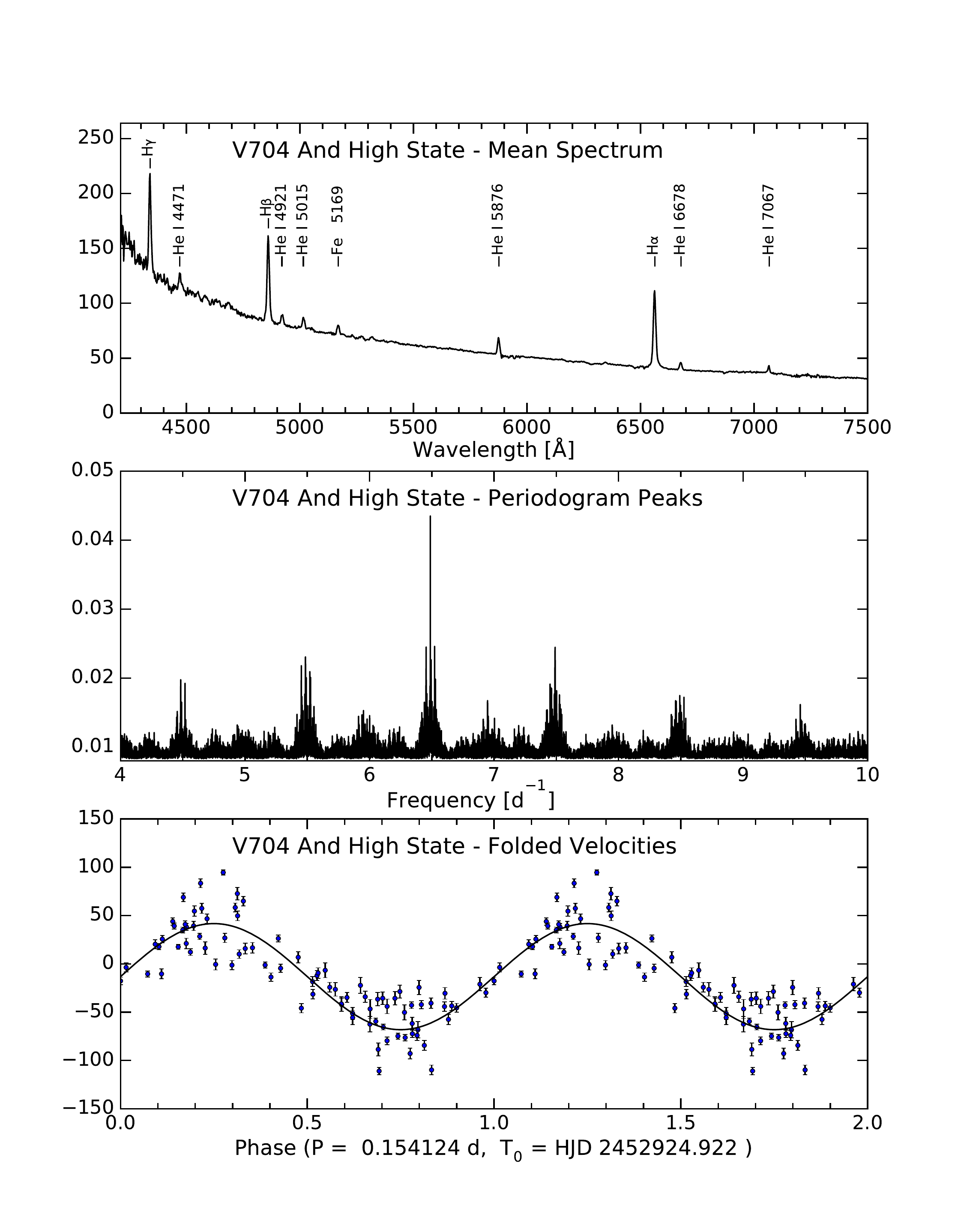}

    \caption{Average spectra, periodograms, and folded velocity curves for
V704~And in the high state. The vertical scales, unlabeled to save space, are
(1) for the spectra, $f_\lambda$ in units of $10^{-16}\textrm{ erg
s}^{-1}\textrm{ cm}^{-2}\textrm{\AA}^{-1}$; (2) for the periodograms,
$1/\chi^2$ (dimensionless); and (3) for the radial velocity curves, barycentric
radial velocity in $\textrm{km s}^{-1}$. The periodogram is labeled with the
word ``peaks'', because the curve shown is formed by joining local maxima in
the full periodogram with straight lines. This suppresses fine-scale ringing
due to the unknown number of cycle counts between runs. The folded velocity
curves all show the same data plotted over two cycles for continuity, and the
best-fit sinusoid is also plotted. The velocities shown are H$\alpha$ emission
velocities. The error bars for the emission lines are computed by propagating
the estimated noise in the spectrum through the measurement, and hence do not
include jitter from line profile variations.}

   \label{fig:V704High}
\end{figure*}

\begin{figure*}[t]
   \centering
   \includegraphics[width=0.75\textwidth]{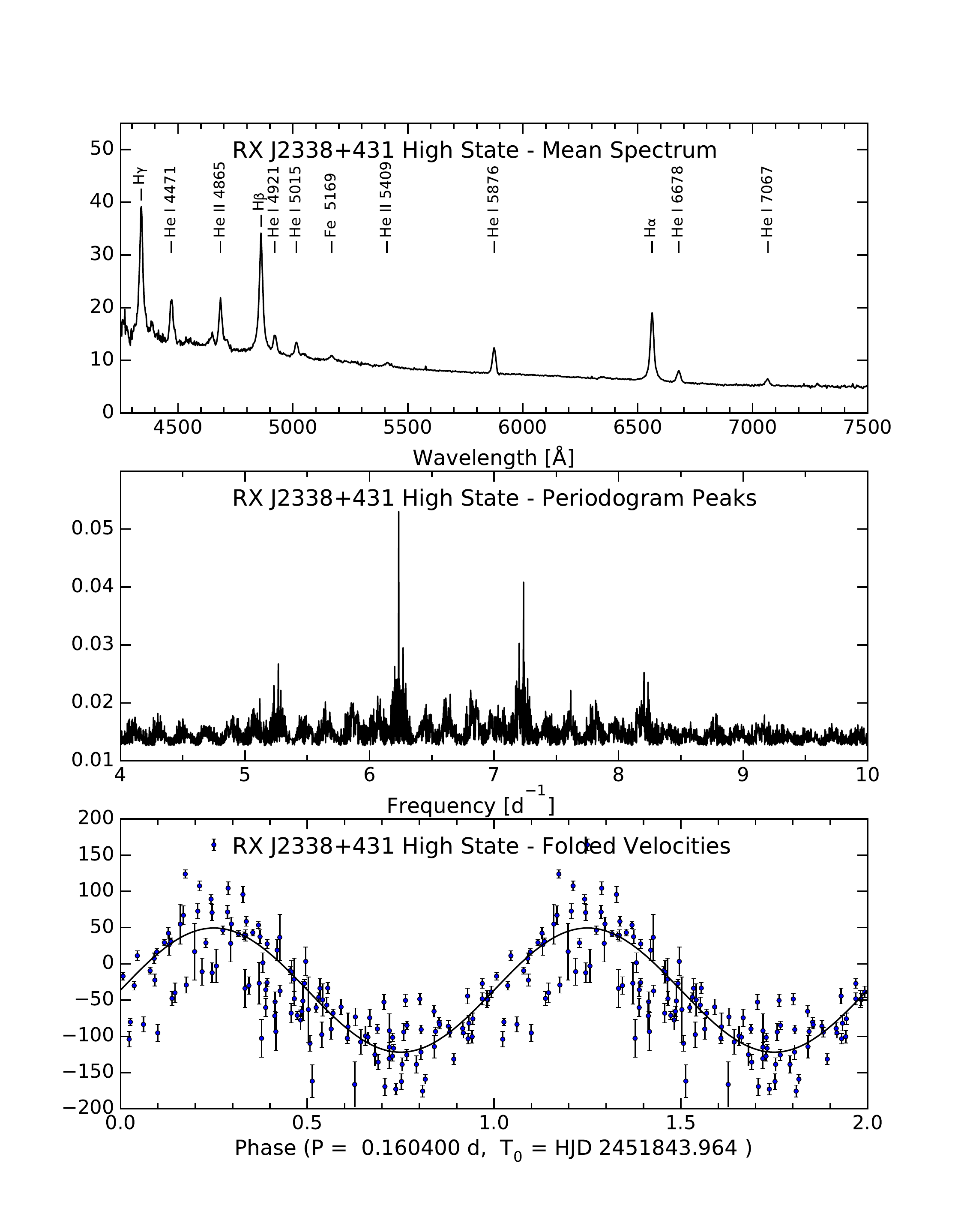} 
      \caption{Similar to Figure~\ref{fig:V704High} but for RX~J2338+431 in the high state.}
   \label{fig:rx2338High}
\end{figure*}

For both the objects, the high-state spectra were the 
most numerous, so we used these to determine $P_{\rm orb}$.
We measured H$\alpha$ emission radial velocities using 
a convolution algorithm \citep{SchneiderYoung1980}, and 
applied a `residual-gram' period search method
described by \citet{Thorstensen2016}.  
The middle panels of Figures~\ref{fig:V704High} \& \ref{fig:rx2338High} show 
periodograms for V704~And and RX~J2338+431 respectively.

Fixing $P_{\rm orb}$ at the values determined from the high-state velocities, we fit 
sinusoids to the high-, mid-, and low-state velocities separately.  
The functional form of the fits was 
\begin{equation}
v(t)=\gamma+K\sin{(2\pi(t-T_0)/P)},
\end{equation}
where $T_0$ is the Heliocentric Julian Date minus 2400000, and $P$ is the period in days (Table~\ref{FitsRV}). 

\begin{deluxetable*}{lllrrrr}[t]
\tablecaption{Fits to Radial Velocities \label{FitsRV}}
\tabletypesize{\scriptsize}
\tablehead{\colhead{Data set}&\colhead{$T_0$\tablenotemark{a}}&\colhead{$P$}&\colhead{$K$}&\colhead{$\gamma$}&\colhead{$N$}&\colhead{$\sigma$\tablenotemark{b}}\\
& & \colhead{(d)}&\colhead{${\rm km\ s}^{-1}$}&\colhead{${\rm km\ s}^{-1}$}& &\colhead{${\rm km\ s}^{-1}$}}
\startdata
V704~And (High) & 52924.2218(27) & 0.154124(3) & 55(4) & -13(3) & 87 & 22\\
V704~And (Intermed.) & 55128.7168(23) & \nodata\tablenotemark{c} & 47(4) & -22(3) & 31 & 19\\
V704~And (Low) & 57430.5220(19) & \nodata\tablenotemark{c} & 27(2) & -6(2) & 47 & 9\\
RX~J2338+431 (High) & 51843.9642(25) & 0.160400(1) & 86(8) & -36(6) & 135 & 35\\
RX~J2338+431 (Low) & 51553.6794(28) & \nodata\tablenotemark{c} & 55(6) & -31(4) &19 & 28\\
\enddata
\tablecomments{Parameters of least-squares sinusoid fits to the radial velocities of the form $v(t)=\gamma+K\sin{(2\pi(t-T_0)/P)}$.}
\tablenotetext{a}{Heliocentric Julian Date minus 2400000. The epoch is chosen to be near the center of the time interval covered by the data, and within one cycle of an actual observation.}
\tablenotetext{b}{RMS residual of the fit.}
\tablenotetext{c}{Period fixed at value derived from high state data.}
\end{deluxetable*}


\section{Results} 

\subsection{V704~And} 

The light curve of V704~And from the Catalina
Real-time Transient Survey (CRTS) \citep{Drake2009},
Figure~\ref{fig:V704rx2338LC} (blue circles), shows repeated dimmings
of $\sim 5$ mag. Additionally,
Figure~\ref{fig:V704rx2338LC} shows good agreement between the V
magnitudes synthesized from our spectra and the observed
photometric light curve. 

\begin{figure*}[t]
   \centering
    \includegraphics[width=0.65\textwidth]{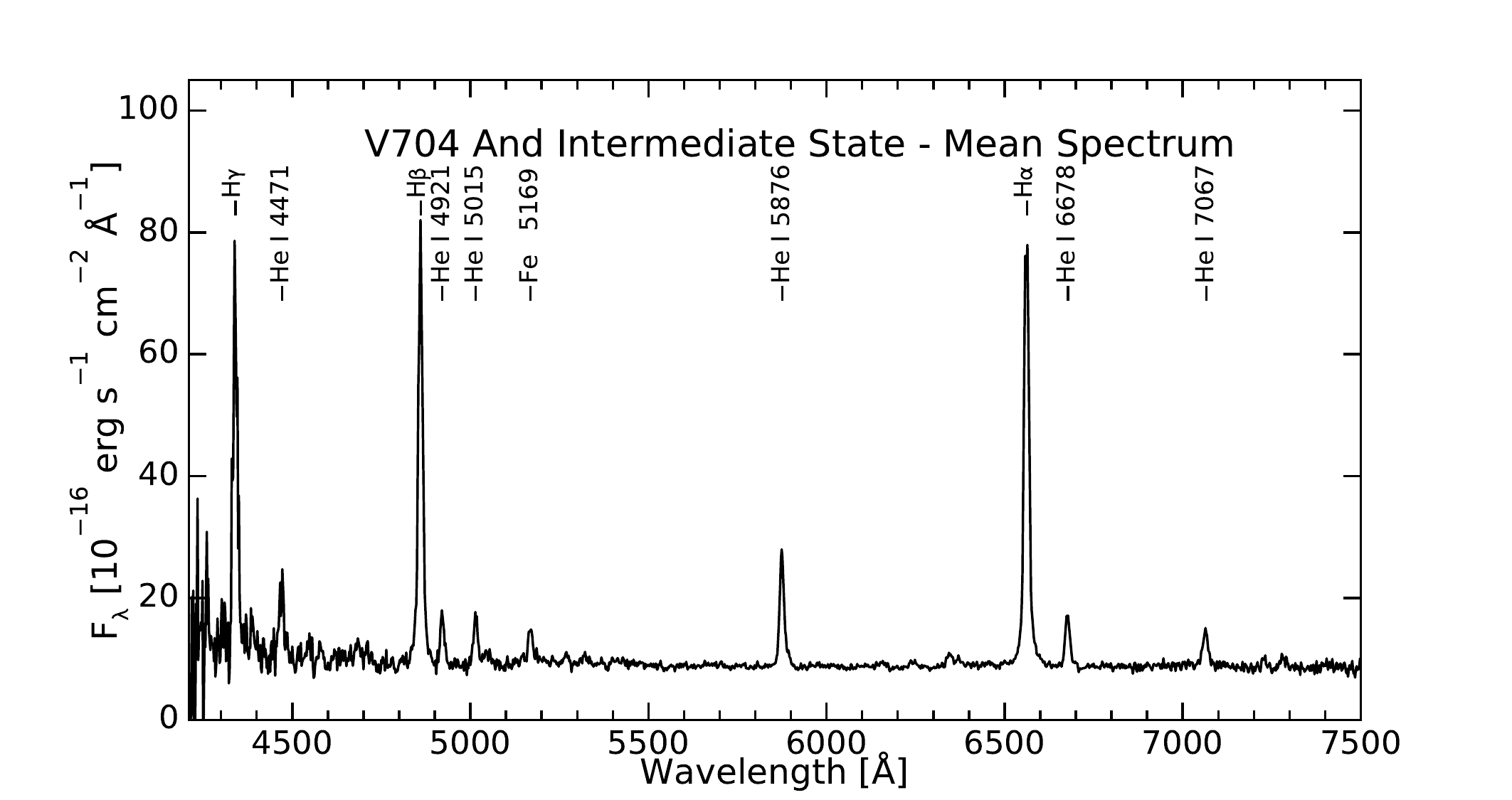}
    \includegraphics[width=0.65\textwidth]{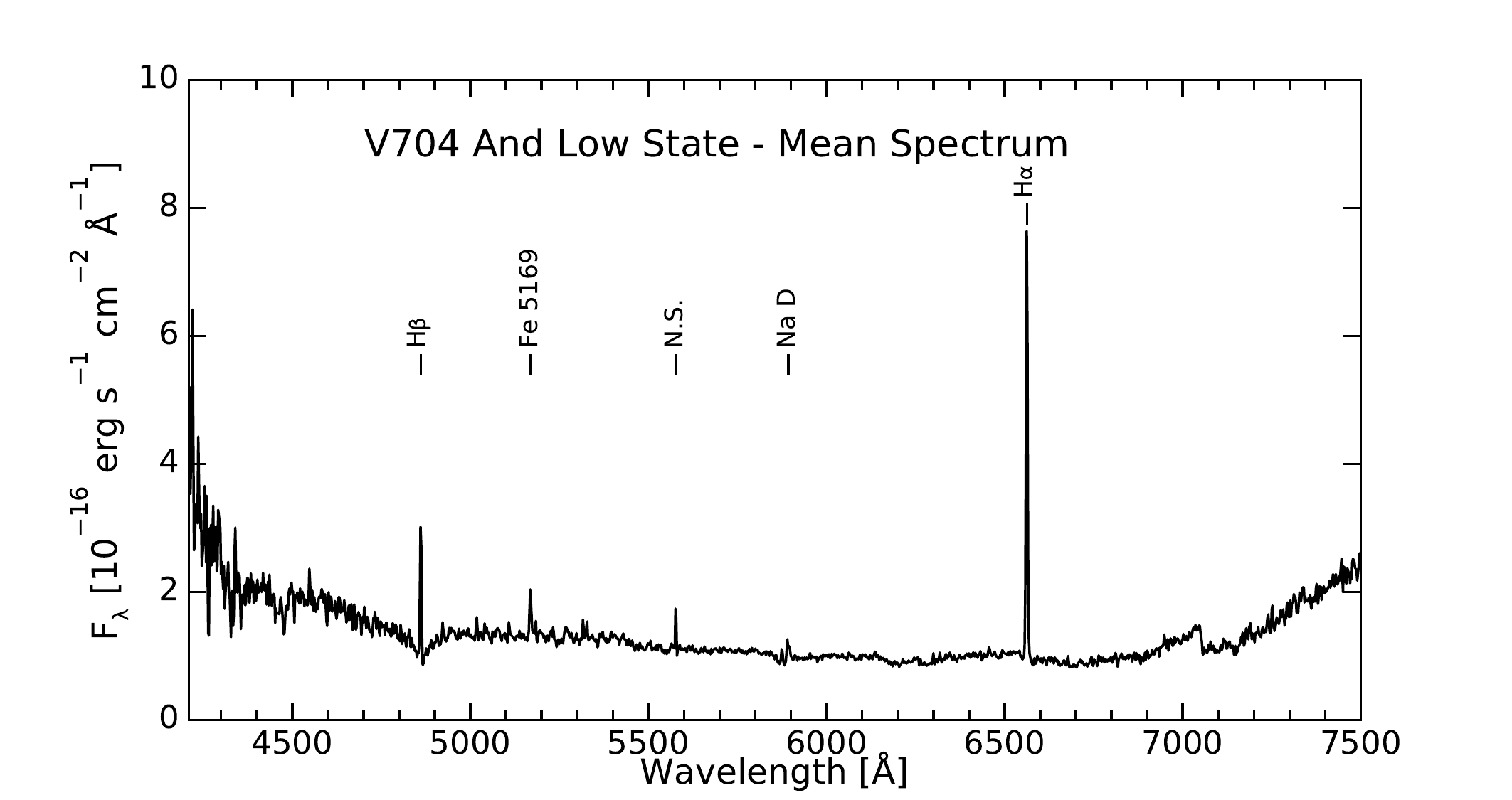}
       \caption{Average spectra for V704~And. {\it Top:} Average intermediate state spectrum from 2008 June. {\it Bottom:} Average low state spectrum from 2016 June. (There was apparently a problem with the flux calibration during the 2016 June observing run, and the continua longward of 7000\AA\ and shortward of 4500\AA\ are not reliable.) Noticeable differences can be seen between the spectra, including overall flux and lack of He I emission during the low state.}
   \label{fig:V704MidLow}
\end{figure*}

The high state spectrum (top panel Figure~\ref{fig:V704High}) shows Balmer
emission lines and emission from He~I $\lambda\lambda$ 4471, 4921, 5015, 5876,
6678, and 7067 as well as weak emission from an Fe blend at $\lambda$5169. 
In the intermediate state (top panel
Figure~\ref{fig:V704MidLow}), the continuum flux has decreased, but the
emission lines have become relatively stronger. In 
the low state, there was both an overall decrease in the continuum
flux and decreased line emission (bottom panel Figure~\ref{fig:V704MidLow}),
with a markedly steeper Balmer decrement.
The \ion{He}{1} and \ion{He}{2} lines are nearly absent, though a small
trace of \ion{He}{1} $\lambda 5876$ may be present.
The wide H$\beta$ absorption in the low state appears to be from the white
dwarf photosphere.

Comparing the spectral appearance to 
the V magnitudes synthesized from the spectral observations,
we found that the high-state spectra corresponded to $12.7 \le V \le 16$,
intermediate state to $16.4 \le V \le 18.3$, and the low state to 
$V = 18.9 \pm 0.2$.  We do not
have enough coverage to say whether the system was brightening
or dimming when the intermediate-state spectra were taken.

The low state spectrum (bottom panel Figure~\ref{fig:V704MidLow}) shows 
features of a late-type secondary star in the far red end,
although most of the spectrum is dominated by the WD continuum. 
To estimate the secondary's spectral type,
we scaled spectra of stars of known spectral type and subtracted them from 
the observed spectrum, with the aim of canceling out the late-type star's 
features and leaving a smooth continuum.  This was complicated by 
an unusually poor flux calibration longward of 7000 \AA\ and shortward of 4500 \AA .
With those caveats, our best estimate for the secondary's spectral type is M3--M4,
and the best scaling gave a secondary contribution equivalent to $V \sim 21$. 
The Gaia Data Release 2 \citep{GaiaDR2}, which was published
as this paper was being revised, gives a parallax of $2.4652 \pm 0.0589$ mas,
which inverts to $406 \pm 10$ pc.  

The periodogram of the high state radial velocities (middle panel
Figure~\ref{fig:V704High}) reveals $P_{orb}=0.151424(3)
\textrm{d}$, or $\sim 3.63$ hr. Periods that differ by one cycle per 9.5 years 
are considerably less likely but not entirely ruled out. 

\begin{figure*}[t]
   \centering
   \includegraphics[scale=0.65]{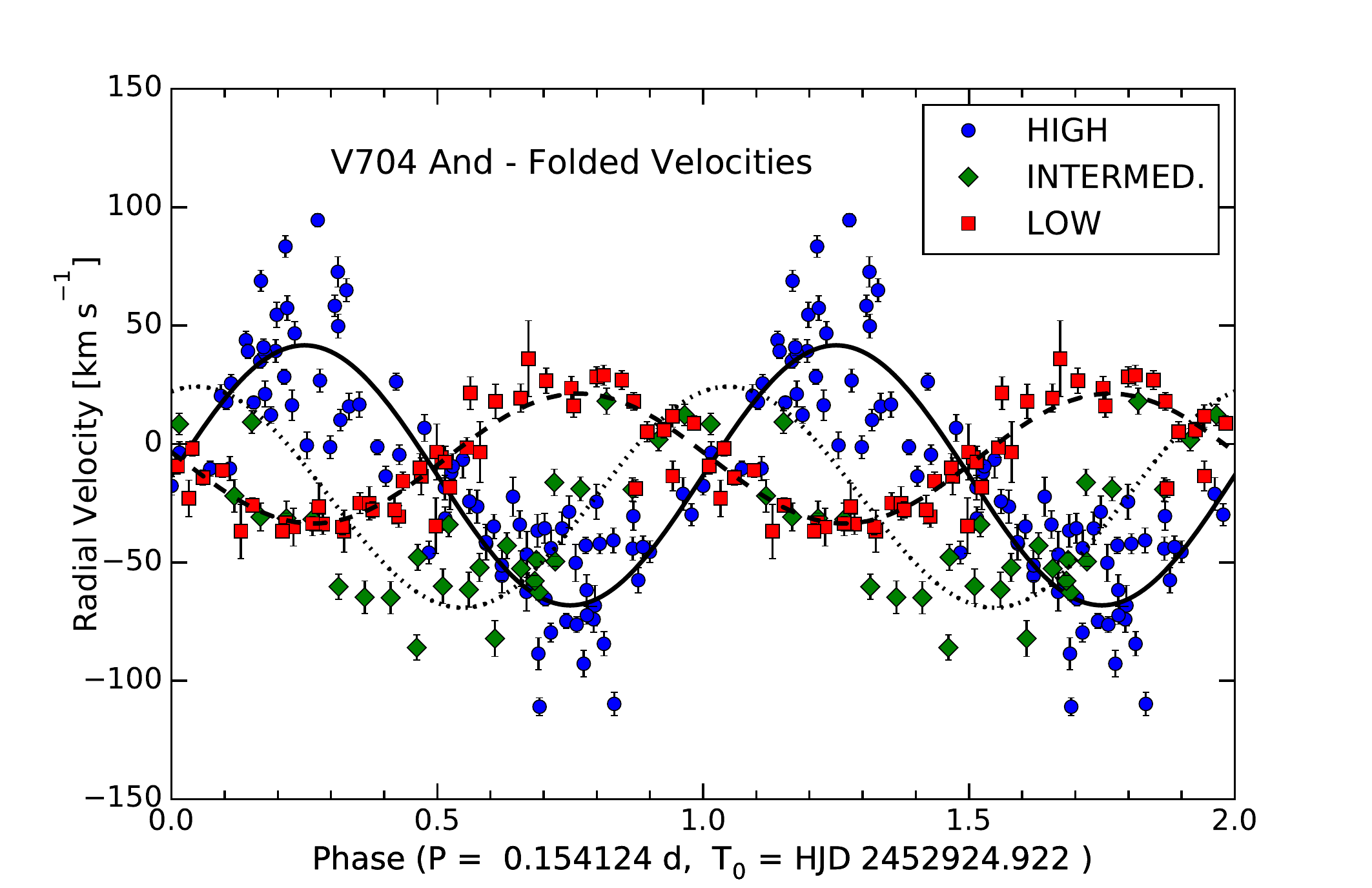} 

   \caption{Folded velocity curves for V704~And in the high (blue circles),
intermediate (green diamonds) and low (red squares) states. The curves all show
the same data plotted over two cycles for continuity, and the best-fit sinusoid
for each state is also plotted. The velocities shown are H$\alpha$ emission
velocities. 
}

   \label{fig:V704Sort}
\end{figure*}

Folding all the velocities together on the orbital period
reveals a clear phase shift between the velocities
from the various states (Figure~\ref{fig:V704Sort}).  The phase shift between the high and
intermediate state is $0.80\pm0.04$ while the shift between the high
and low state is $0.52\pm0.07$.  This is not unprecedented; a phase shift 
between the high and low state of approximately 180 degrees is seen 
in TT Ari \citep{Shafter1985}.

\subsection{RX~J2338+431} 

The light curve from direct imaging shows
RX~J2338+431 changing states from high to low in 3 months between 
1999 October and 2000 January (Figure~\ref{fig:V704rx2338LC}). The
2000 October observations show it back in the high state.

\begin{figure*}[t]
   \centering
   \includegraphics[width=0.65\textwidth]{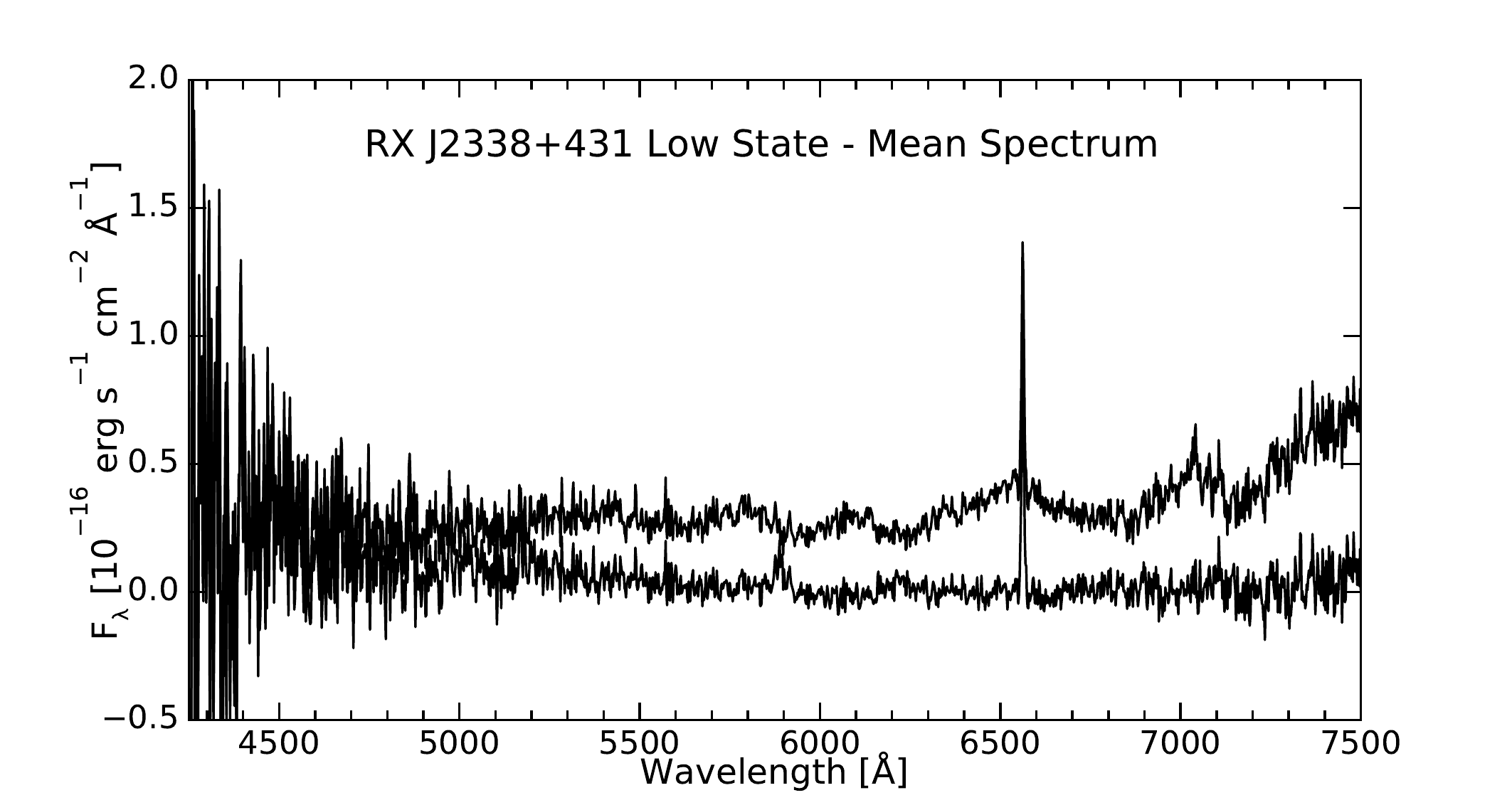} 

   \caption{Average spectrum RX~J2338+431 in the low state from 2000 January.
The top trace shows the average spectrum, and the bottom shows the average
spectrum after subtraction of a scaled M2.5-type spectrum.}

   \label{fig:rx2338Low}
\end{figure*}

In the high state, RX~J2338+431 shows Balmer emission lines and
emission from \ion{He}{1} $\lambda\lambda$ 4471, 4921, 5015, 5876, 6678, and 7067,
\ion{He}{2} $\lambda\lambda$ 4685, 5409, as well as weak emission from an Fe blend at
$\lambda 5169$ (top panel Figure~\ref{fig:rx2338High}). During the low state
(Figure~\ref{fig:rx2338Low}) the overall flux is much lower, and only narrow
H$\alpha$ emission is present. H$\beta$ is not detected in emission or
absorption, probably due to the poor signal-to-noise at $\lambda<5000$\AA.  

The low state average spectrum clearly shows a late-type contribution
from the secondary star (Figure \ref{fig:rx2338Low}).  We 
estimated the spectral type and flux of the secondary by subtraction,
in a manner similar to that used for V704 And.  For the type, we 
estimate  M2.5$\pm$0.5, and the normalization corresponds to V $\sim20.6$. 
Following the techniques described in \citet{ThorstensenHalpern2013}, we used the 
observed flux and spectral type of the secondary, and a constraint on the 
secondary's radius (derived from the orbital period and a plausible range for its mass),
to estimate a distance of $890\pm200{\rm\ pc}$.  
The Gaia Data Release 2 \citep{GaiaDR2} parallax is $1.0742 \pm 0.0722$ mas,
corresponding to $d = 930 \pm 60$ pc, in excellent agreement 
with the photometric estimate.  

\begin{figure*}[t]
   \centering
   \includegraphics[scale=0.65]{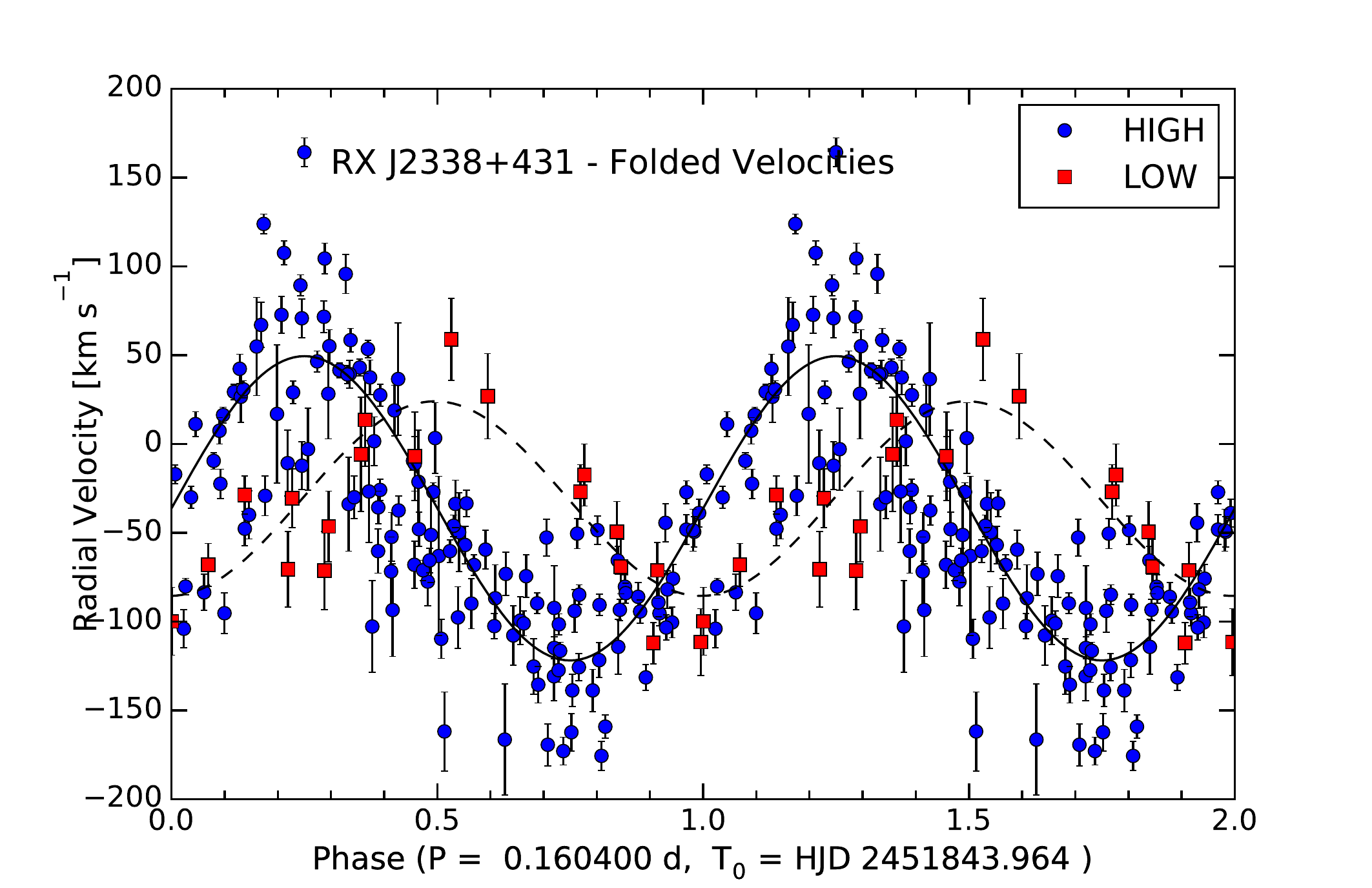} 

   \caption{Folded H$\alpha$ emission velocity curves for RX~J2338+431 in the high (blue circles)
and low (red squares) states. The same data are shown repeated over
two cycles for continuity, and the best-fit sinusoid for each state is also
plotted.}

   \label{fig:rx2338Sort}
\end{figure*}

The periodogram of the high state H$\alpha$ radial velocities (middle panel
Figure~\ref{fig:rx2338High}) reveals $P_{orb}=0.160400(1)\textrm{
d}$, or $\sim 3.13$ hr. The likelihood of a cycle count error on day-to-day or run-to-run
timescales is very small. When the radial velocities from high and low states
are examined together (Figure~\ref{fig:rx2338Sort}) a phase shift of
$0.25\pm0.03$ is observed.

\section{Discussion}

In many low state systems, there appears to be no evidence for emission from
the disk. This is also the case for V704~And and RX~J2338+431, as the emission
lines that are seen are quite narrow, 
which is not consistent with the rotation broadening expected from 
a disk. 

The origin of the narrow emission lines seen in the low states of VY Scl stars
is poorly understood.  \citet{Warner1995} suggested that they could arise from irradiation 
of the secondary by the WD.  If that were the case, the
emission strength would be expected to vary with orbital phase, as the 
irradiated face rotates in and out of view (\citealt{tcmb78} present
an early and clear-cut example of this phenomenon).
In both of the systems studied here, the EW of the low-state H$\alpha$
does {\it not} vary systematically with phase, which would suggest
that the emission comes from the whole surface, as from
a chromosphere.  However, the observed velocity amplitudes $K$
are much smaller than would be expected in this scenario; for 
plausible component masses, the observed $K$-velocities would require 
orbital inclinations $i$ less than $\sim 20\arcdeg$. For randomly oriented
orbits, the probability of $i < 20\arcdeg$ is only $0.06$, so the 
probability of finding two such systems at random is $0.06^2$ or 
0.0036.  

Several other VY Scl stars also show similarly puzzling
behavior in their low-state H$\alpha$ emission.
\citet{schmidtobreick12} present time-resolved VLT spectroscopy
of the VY Scl star BB Dor in a low state.  Their data are 
more extensive and have higher spectral resolution than ours,
and show NaD emission lines that must arise on the 
secondary -- they were able to trace the secondary's 
motion by following a TiO band head, and the NaD emission
was modulated similarly.  They also detect narrow
H$\alpha$ emission.  Because they had traced the secondary's motion 
independently, they established that the H$\alpha$ moves 
approximately in phase with the 
secondary, but with a significantly lower amplitude.  
(They also detect remarkable `satellite emission' with velocities 
roughly consistent with gas orbiting near the L4 and L5 points; 
these do not appear in our data, perhaps because of our
relatively modest signal-to-noise).
Curiously, as in our data, they do not see an orbital modulation in the emission
strength, as would be expected if the line originated at the heated face
of the normal star. \citet{RG2012}, in a study of the eclipsing 
VY Scl star HS0220+0603, similarly find that the modulation of the 
H$\alpha$ emission velocity has an amplitude $K \sim 100$ km s$^{-1}$, 
less than half that of the photospheric absorption at CaII (at $\lambda\lambda 8183, 8194$).
They suggest that the low H$\alpha$ velocity amplitude reflects an origin 
mostly on the facing hemisphere of the secondary. 
However, the H$\alpha$ line remains visible around
zero (eclipse) phase, so the emission cannot arise entirely from the
hemisphere facing the white dwarf.  The sharpness,
low velocity amplitude, and lack of orbital modulation of the 
H$\alpha$ line once again presents a puzzle.



The presence of relatively strong \ion{He}{2} in the high state spectrum of
RX~J2338+431 suggests a magnetic system. X-ray selected CVs also tend to be
disproportionately magnetic \citep{ThorstensenHalpern2013}.  The absence of
dwarf nova outbursts during VY Scl low states provide indirect evidence
for a magnetic field; \citet{hameury02} point out that according to
disk instability theory, VY Scl stars in low states should show outbursts 
unless their disks are disrupted, as by a magnetic field  
\footnote{
\citet{KingCannizzo1998} had suggested earlier that dwarf-nova outbursts
could be suppressed if the disk were irradiated by an unusually hot
white dwarf, but \citet{hameury02} found that suggestion to fail
quantitatively unless the white dwarf were $\sim 0.4$ M$_{\odot}$ or
less (and hence of unusually large radius); this is well below typical
white dwarf masses in CVs.}.
Time series photometry of RX~J2338+431 might detect the regular
pulsations characteristic of magnetic CVs (i.e. DQ~Her stars or
IPs).  

\section{Summary} We present photometric and spectroscopic observations of the
cataclysmic variable systems V704~And and RX~J2338+431. To our knowledge the
optical identification of RX~J2338+431 has previously appeared only in the \citet{downes01} catalog.
We also describe a
method to discriminate the different states of a VY~Scl star system based upon
the EW and FWHM of the H$\alpha$ line. Key findings about the two systems
presented in this paper are as follows:

-- Through spectroscopic observations spanning 15 years, we confirmed the
VY~Scl type classification of V704~And.

-- Through photometry and spectroscopy, we identify RX~J2338+431 as a CV of the
VY~Scl subclass. 

-- We determine the orbital periods to be
$P_{orb}=0.151424(3) \textrm{ d}$ and $P_{orb}=0.160400(1)\textrm{ d}$ for
V704~And and RX~J2338+431, respectively, which fall in the period range  
typical of VY~Scl novalikes.

-- Phase shifts of $0.52\pm0.07$ and $0.25\pm0.03$ for V704~And and
RX~J2338+431, respectively, were found between the high and low states of the
systems, indicating a change in the source of the emission line. 

-- Using the low state spectrum of RX~J2338+431, we determine the secondary's
spectral type to be M2.5$\pm$0.5, and estimate a photometric distance of 
$890 \pm 200{\rm\ pc}$, in good agreement with the nominally more 
accurate Gaia DR2 parallax.

\acknowledgments

We gratefully acknowledge support from NSF (AST 99-87334, AST 03-07413, AST
0708810 and AST-1008217), and thank the MDM staff for observatory support.

We acknowledge with thanks the variable star observations from the AAVSO
International Database contributed by observers worldwide and used in this
research. 

The CSS survey is funded by the National Aeronautics and Space Administration
under Grant No. NNG05GF22G issued through the Science Mission Directorate
Near-Earth Objects Observations Program.  The CRTS survey is supported by the
U.S.~National Science Foundation under grants AST-0909182 and AST-1313422.

This work has made use of data from the European Space Agency (ESA) mission
{\it Gaia} (\url{http://www.cosmos.esa.int/gaia}), processed by the {\it Gaia}
Data Processing and Analysis Consortium (DPAC,
\url{http://www.cosmos.esa.int/web/gaia/dpac/consortium}). Funding for the DPAC
has been provided by national institutions, in particular the institutions
participating in the {\it Gaia} Multilateral Agreement.

Funding for the Sloan Digital Sky Survey IV has been provided by the Alfred P.
Sloan Foundation, the U.S. Department of Energy Office of Science, and the
Participating Institutions. SDSS-IV acknowledges support and resources from the
Center for High-Performance Computing at the University of Utah. The SDSS web
site is \url{www.sdss.org}. A full listing of institutions can be found at
\url{www.sdss.org}.

Finally, we thank the anonymous referee for a constructive and useful
report.


\facility{Hiltner (Modspec+Echelle)}, \facility{McGraw-Hill (Modspec+Echelle)}, \facility{AAVSO}

{}

\end{document}